\documentclass[prl,aps,twocolumn,floats]{revtex4} \input epsf
\usepackage{bm}
\begin{document}
\title{Chemosensing in microorganisms to practical biosensors}
\author{Surya K. Ghosh}
\email{surya@phy.iitb.ac.in}
\affiliation{Department of Physics, Indian Institute of Technology, 
Bombay, Powai, Mumbai-400 076, India. }
\author{Tapanendu Kundu}
\affiliation{Department of Physics, Indian Institute of Technology, 
Bombay, Powai, Mumbai-400 076, India. }

\author{Anirban Sain}
\affiliation{Department of Physics, Indian Institute of Technology, 
Bombay, Powai, Mumbai-400 076, India. }

\begin{abstract}
Microorganisms like bacteria can sense concentration of chemo-attractants in its 
medium very accurately. They achieve this through interaction between the receptors 
on their cell surface and the chemo-attractant molecules (like sugar). But the 
physical processes like diffusion set some limits on the accuracy
of detection which was discussed by Berg and Purcell in the late seventies. We 
have a re-look at their work in order to assess what insight it may offer towards
making efficient, practical biosensors. We model the functioning of a typical 
biosensor as a reaction-diffusion process in a confined geometry. Using available
data first we characterize the system by estimating the kinetic constants for the
binding/unbinding reactions between the chemo-attractants and the receptors. Then 
we compute the binding flux for this system which Berg and Purcell had discussed. 
But unlike in microorganisms where the interval between successive measurements 
determines the efficiency of the nutrient searching process, it turns out that 
biosensors depend on long time properties like signal saturation time which we
study in detail. We also develop a mean field description of the kinetics of the 
system.
\end{abstract}
\maketitle

Berg and Purcell(BP), in their pioneering article \cite{BP} on ``physics of chemoreception" 
had considered how a micro-organism could sense concentration of a chemo-attractant
molecule (say, X) in its surrounding media. They assumed the organism to be a sphere 
of radius $a$, immersed in an unbounded liquid medium and $\rho_0$ be the far field 
concentration of X. A simple example could be a bacterium in a dilute sugar (X) solution
of local density $\rho$. The X molecules diffuse and bind to the surface of 
the sphere which is assumed to be a perfect sink for X. They solved Diffusion equation, 
$\partial _t \rho=D\nabla ^2 \rho$, in the steady state, using spherical co-ordinates
centered at the sphere. Using the boundary conditions $\rho(r=a)=0$ (i.e., fully 
absorbing surface) and $\rho(r=\infty)=\rho_0$ (at far field) they obtained the 
steady state influx of X molecules ($J$) integrated over the spherical surface 
to be $J=4\pi aD \rho_0$.

We will now briefly introduce a typical biosensor and discuss the applicability of the
above ideas. A biosensor is designed to detect traces of specific biochemicals 
present in a carrier medium. It can detect, for example, {\em E. coli} in drinking water
\cite{geng}, hepatitis B surface antigen present in human serum\cite{sheng} or  
pollutants in air \cite{sara}. The last decade has seen proliferation of such 
biosensors \cite{wilason,bondi,sii} in day to day use, mainly due to their, 
(a) quick response time\cite{ligler}, (b) sensitivity to minute amount of 
biomolecules \cite{geng}.
The particular type of biosensors we discuss here are optics based chemical
sensors which converts chemical reactions between GaHIgG (X) and HIgG (receptor) 
molecules into optical signal which is then detected using fiber-optics technology. 
In this sensor, an optical fiber of radius $R_i$ runs along the axis of a
cylindrical chamber of radius $R_o$. 
The fluid containing
the antigen (X) is injected into the annular space between the fiber and the chamber.
The surface of the fiber is functionalized by putting a certain surface density 
($\sigma_0$) of antibodies(receptors) on it which serve as the binding targets 
for the antigen molecules.  Antigens bind to the receptors on the surface of the 
fiber and absorb evanescent waves generated by the light carrying fiber. This 
results in loss of intensity carried by the fiber. For our purpose here, the 
absorbance($\cal A$) of the evanescent waves \cite{ruddy} is proportional to the 
total bound antigen $\int \sigma dA$ on the fiber surface, where $\sigma$ is 
the surface density of bound antigens.

Such a system can be described, at the continuum level, by reaction diffusion 
equations \cite{huang}. 
The X molecules bind to the receptors on the fiber surface with a rate $\omega_b$ 
and surface bound X molecules can also unbind at a rate $\omega_u$, typically
much smaller than the binding rate. The values of the kinetic coefficients 
$\omega_b$ and $\omega_u$ are unknown a priori which we will determine from
experimental data.  The bulk concentration of X is $\rho$, the surface 
concentration of receptor-bound X molecules is $\sigma$ and the surface 
concentration of receptors be $\sigma_0$.
Dynamics of $\rho$ follows
\begin{equation}
\frac{\partial \rho}{\partial t} =  D \nabla^2 \rho -\delta(r-R_i)[\rho (\sigma_{o} - \sigma)  
\omega_b  - \omega_u \sigma] 
\end{equation}
We will use cylindrical polar coordinate frame where $\rho=\rho(r,\phi,z,t)$
and $\sigma=\sigma(\phi,z,t)$ with $R_o>r>R_i$. The second term on the right hand
side represents surface reactions at $r = R_{i}$.  The first term in the square 
bracket describes binding and the second term represents unbinding. Dynamics of 
$\sigma$ follows
\begin{equation}
\frac{\partial \sigma}{\partial t} = \rho (\sigma_{o} - \sigma)  \omega_b  - \omega_u \sigma. 
\label{surfDim}
\end{equation}
Here $\rho$ is the bulk density in the immediate vicinity of the surface.
These equations can be nondimensionalized. We rescale the bulk and 
the surface densities as $\tilde{\rho}=\frac{\rho}{\rho_0}$, 
and $\tilde{\sigma}=\frac{\sigma}{\sigma_{0}}$;
the space and time variables as $\tilde r=r/R_i, \tilde z =z/L$ 
and $\tau=\frac{t D}{R_{i}^{2}}$. In terms of dimensionless parameters
$\tilde{\omega_{b}}=\omega_{b} \beta,\;\tilde{\omega_{u}}=
\frac{\omega_{u}}{\rho_{0}} \beta$ and $\gamma=\frac{\sigma_{0}}{\rho_{0}R_i}$, 
where $\beta=\frac{R_{i}^{2}}{D}\rho_{0}$, the equations are
\begin{eqnarray}
\frac{\partial \tilde{\rho}}{\partial \tau} &=&   {\tilde{\nabla}^2 \tilde{\rho}} -\delta(\tilde{r}-1)
\gamma\left[ \tilde{\rho} 
(1 - \tilde{\sigma})  \tilde{\omega_b}  - \tilde{\omega_u} \tilde{\sigma} \right]
\label{diffsnNonDim}\\
\frac{\partial \tilde{\sigma}}{\partial \tau} &=& \tilde{\rho} (1 - \tilde{\sigma})  
\tilde{\omega_b}  - \tilde{\omega_u} \tilde{\sigma}.
\label{surfNonDim}
\end{eqnarray}

\begin{figure}[htbp]
\epsfxsize=7.5cm
\centerline{\epsfbox{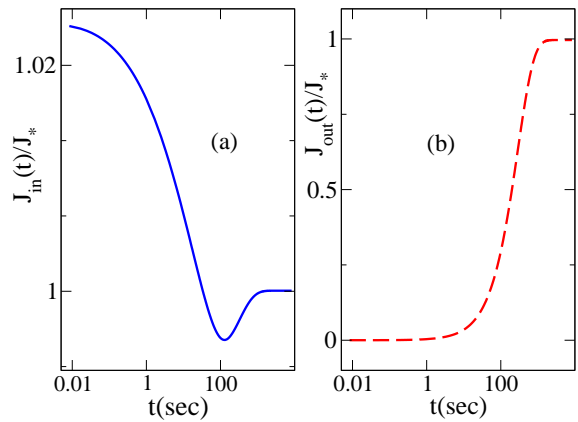}}
\caption{
Semi-log plot of binding ($J_{in}$) and unbinding ($J_{out}$) fluxes (both scaled with the steady state
value $J_{*}$) as a function of time, at fixed $D$ and $\rho_0 (=0.001mg/ml)$. 
Values of $\omega_{b}$ and $\omega_{u}$ are same as those found through Fig.\ref{fig.absVsTime}
(to be discussed later). $J_{in}$ and $J_{out}$ are plotted separately because their scales
of variations are very different, unlike that in Fig.\ref{fig.JHighWb}} 
\label{fig.JLowWb}
\end{figure}

\begin{figure}
\vspace{0.0cm}
\epsfxsize=6.5cm
\begin{center}
\centerline{\epsfbox{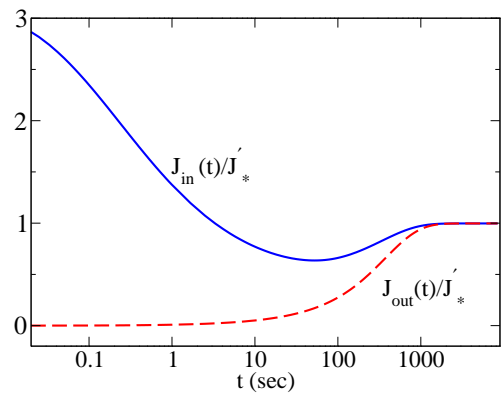}}
\caption{
Semi-log plot of binding ($J_{in}$) and unbinding ($J_{out}$) fluxes (both scaled with the steady state
value $J_{*}^{'}$) as a function of time. Here we used the same $\rho_{0}$, $D$ and $\omega_{u}$ 
as in Fig.\ref{fig.JLowWb} while $\omega_{b}$ was increased $100$ times to reach a steady 
state value $J_{*}^{'}$ comparable to $J_{BP}$. We explain later why this comparison
is not very useful due to difference in the boundary conditions in the two problems.}
\label{fig.JHighWb}
\end{center}
\end{figure}

Superficially the spherical surface of BP is replaced in our biosensor by a cylindrical
surface but the big difference is that our system is confined and the total number
of antigens is fixed. Thus the steady state here corresponds to a state of dynamic
equilibrium when the binding and unbinding at the fiber surface balance each other
making both the bulk  and surface concentrations constant is time. Note that although
the surface concentration becomes static the steady state binding ($J_{in}$) and 
unbinding ($J_{out}$) fluxes, individually are not zero at the surface (see 
Fig\ref{fig.JLowWb},\ref{fig.JHighWb} obtained at different values of $\omega_b$) 
and at large time $J_{in}=J_{out}=J_*$. In BP's
case, with perfectly absorbing surface $J_{in}$ is given by the surface integral of
the diffusional current $-D\int \vec \nabla \rho dA$ onto the absorbing surface 
(of area $A$). But for our sensor with finite binding constant the influx 
$J_{in}=\int dA\rho (\sigma_{o} - \sigma)  \omega_b$ is computed
by integrating the binding term (on the right hand side of Eq.2) over the area $A$ of
the fiber surface, and similarly the outflux $J_{out}=\int dA\omega_u\sigma$ is computed 
from the unbinding term. We consider the initial condition where at $t=0$ the system
is filled up with a fluid carrying an uniform concentration of X. 
At $t=0$ the influx is nonzero as the concentration of X-molecules in the
vicinity  of the fiber surface ($\rho_s$) is non-zero. On the other hand the outflux 
is zero at $t=0$ because there are no bound X-molecules at the beginning. 
BP has considered a perfectly absorbing surface which can theoretically be attained in 
the limit $\sigma_0\rightarrow \infty$ and $\omega_u\rightarrow 0$. Note that, in Eq.1 
and 2, when $\sigma_0\gg \sigma$ the binding term reduces to $\rho_s\sigma_0\omega_b$ 
and it appears that we do not need $\omega_b$ to be infinity in addition.
But practically $\sigma_0$ is bounded due to the  finite size of the receptors.
BP had approximated the receptors to occupy a small area with radius $s\sim 10 \AA$. 
For our sensor it amounts to about $10^{13}$ receptors covering the whole fiber surface. 
We used this value as the maximum coverage $\sigma_0^{max}$ for Fig.\ref{fig.JLowWb},
\ref{fig.JHighWb}. To compare with BP's case we will focus on the binding flux of X only.

First we adapt BP's general expression \cite{BP} for the steady flux
$J=4\pi aD \rho_0$ to our cylindrical geometry. BP had shown that this flux can
be calculated for any shape by mapping the steady Diffusion equation 
$\nabla ^2 \rho=0$ to the Poisson equation for potential $\nabla ^2 \phi=0$, 
in charge free space. It can be shown that generally $J=4\pi CD \rho_0$ where 
$C$ is the capacitance of a conductor with free charge $Q$ on its surface. 
Specifically, $C=Q/\phi_{\infty}$, where $\phi_{\infty}$ is the potential 
difference between the conductor and infinity. For the sensor the cylindrical 
fiber is the absorbing surface. With a radius $R_i=0.1mm$ and length $L=50mm$ (i.e.,
aspect ration $500$) it is as good as an one dimensional line. For a line
charge density $\lambda$, extending from $x=-L/2$ to $L/2$ the expression for 
the potential $\phi(z)$ along the perpendicular bisector, $z$ distance away 
from the center of the line charge, is
$\phi(z)=k\frac{\lambda}{\epsilon_0}\ln\frac{\sqrt{1+(z/L)^2}+1}{\sqrt{1+(z/L)^2}-1}$,
where $k=1/4\pi\epsilon_0$ is the Coulomb force constant.
Using the approximation $R_i/L\ll 1$ we get, $\phi(R_i)\simeq\frac{2kQ}{L}\ln(2L/R_i)$
and $C\simeq \frac{L}{2}\ln(2L/R_i)$. Thus $J_{BP}\simeq 2\pi D L\rho_0\ln(2L/R_i)$.
\vspace{0.5cm}

\begin{figure}[htb]
\begin{center}
\epsfxsize=7.5cm
\centerline{\epsfbox{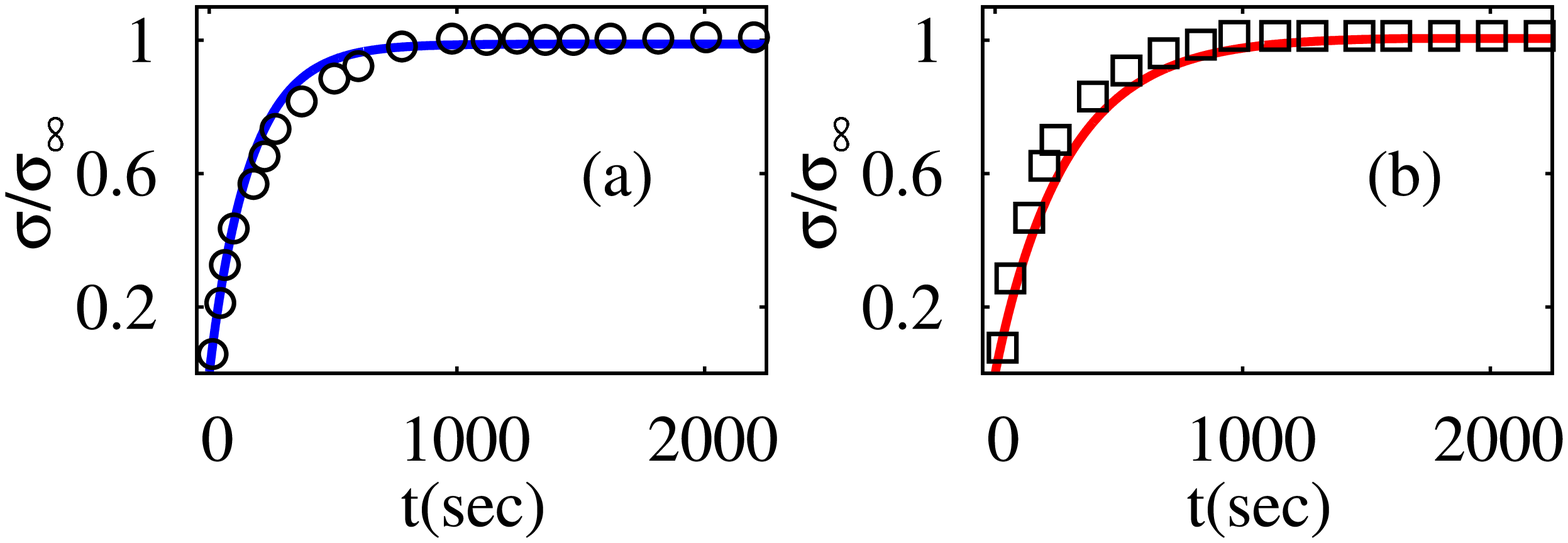}}
\caption{ 
Surface density of bound antigens $\sigma (t)$ versus time. 
y-axis is scaled with the saturation value $\sigma_{\infty}=\sigma(t\rightarrow\infty)$
since we do not know the proportionality constant connecting experimentally measured
absorbance $\cal A$ and $\sigma$.
The symbols represent experimental data. In (a) $\rho_0=0.001 mg/ml$ and in (b) 
$\rho_0=0.1 mg/ml$. The solid lines are  from our numerical integration of Eq.3,4. 
For (a) we chose $\sigma_{0}=0.08\mu g/mm^{2}$ i.e., the maximum possible surface coverage. 
It turns out that a reasonably good match with the two experimental $\sigma (t)$ 
profiles and with $\sigma_{\infty}^a/\sigma_{\infty}^b$ were obtained for 
$\omega_b=0.75 \times 10^{-5} \mu m^{3}/sec, \omega_u=0.35 \times10^{-2} /sec$ and 
$\sigma_0^b=0.014\mu g/mm^{2}$, i.e., about $5.5$ times less than the maximal
coverage.  
}
\label{fig.absVsTime}
\end{center}
\end{figure}

To compute the binding flux $J_{in}$ we have to numerically time evolve the dynamical equations 
(Eq.3,4). First, to get realistic values for the kinetic coefficients $\omega_{b}$ and
$\omega_{u}$ we use experimental data on surface adsorption $\sigma(t)$ versus time from 
Ref\cite{tkundu}, obtained at two widely different initial bulk densities a) $\rho_0=0.001mg/ml$ 
and b) $\rho_0=0.1mg/ml$, and possibly at different surface density of receptors. 
Note that the non dimensionalized equations Eq.\ref{diffsnNonDim},\ref{surfNonDim} do not 
explicitly scale with antigen (X) density $\rho_{0}$ and therefore these data sets can 
be treated as independent. Despite the wide difference in $\rho_0$ the saturation times ($\tau_0$) 
in the two cases were similar (the symbols in Fig.\ref{fig.absVsTime}). This could be
rationalized by noting that, in case-b the fiber was soaked in the receptor solution for two 
hours while for case-a it was soaked for a very long time (about $16$ hours). From this
information we inferred that in case-a $\sigma_0^a=\sigma_0^{max}$ while for case-b 
$\sigma_0^b<\sigma_0^{max}$. We choose $D=10^{-5} cm^2/sec$ typical of diffusion of small 
molecules in water \cite{pastor,kim} (BP also took the same $D$ 
for their estimates). We had to determine $\omega_b, \omega_u$ and $\sigma_0^b$ 
by matching our numerical results (from Eq.3,4) with the temporal profiles of $\sigma(t)$ 
and the ratio $\sigma_{\infty}^a/\sigma_{\infty}^b$. We converged to $\omega_{b}=
0.75 \times 10^{-5} \mu m^{3}/sec$,
$\omega_{u}= 0.35 \times10^{-2} /sec$ and $\sigma_{0}^b=0.014\mu g/mm^{2}$.
These numbers for $\omega_u,\omega_b$ appear reasonable when compared to the 
reaction-diffusion processes on bacterial membrane \cite{huang}.

As mentioned earlier we used cylindrical 
polar coordinate system to discretize the space.  Uniform binning was
used along $z$ and $\phi$; while $r$ coordinate was binned non uniformly such
that the volume of each bin ($rdrd\phi dz$) remains constant. Reflecting boundary
condition was used at the walls of the cylindrical chamber, by ensuring zero
currents at the boundaries. We used an uniform distribution of X molecules in the 
bulk as our initial condition, i.e., $\rho(t=0)=\rho_0$ and $\sigma(t=0)=0$.

We then compute the flux $J_{in}(t)=\int dA\rho (\sigma_{0} - \sigma)  \omega_b$, 
which is the binding term in the right hand side of Eq.2, 
integrated over the cylindrical fiber surface, as a function of time. Interestingly,
$J_{in}(t)$ goes through a minima before it saturates to $J_*$ (see Fig.1,2). We will 
explain the origin of this non-monotonic behavior later when we study the dynamics in 
detail. In Fig.1 the steady state flux $J_*$ is much lower than $J_{BP}$, while in 
Fig.2 it is comparable. But $J_*$ and $J_{BP}$ depends on different set of parameter
values. Both of them are steady state properties, but $J_*$ depends on $\omega_b,
\omega_u,\sigma_0$ and $\rho_0$ while $J_{BP}$ depends on $D$ and $\rho_0$. This 
difference arise from the difference in the boundary conditions of a confined 
versus an unbounded system. Therefore the comparison is not fare. 
$J_*$ can be calculated by setting the left hand side of Eq.4 to zero and using mass 
conservation, which will be discussed later. 

So far we had implicitly assumed that the microorganism can sense the ambient 
$\rho_0$ by measuring the influx ($J$) of X molecules. But BP had also considered 
the realistic possibility that they can infer $\rho_0$ by measuring the state of occupation 
of its surface receptors i.e., density of receptors that are bound to X molecules. 
In fact this is the recipe which most practical biosensors employ. For example, in 
our particular sensor $\sigma(t)$ decides the intensity of optical adsorption. 
In BP's theory a bacteria can sense its $\sigma(t)$ in response to local $\rho_0$ and decide to
move towards or away from the chemo-attractant or the chemo-repellent, respectively. 
But for a static biosensor $\sigma(t)$ can only increase towards a saturation. Since 
a system takes some time to attain saturation, this measurement process is inherently
slow compared to the measurement of instantaneous flux. On the other hand measurement 
of any instantaneous variable is prone to fluctuation error where as long time observables 
like $\sigma(t\rightarrow\infty)$ are more dependable. So the challenge is either to 
reduce the saturation (waiting time) time or choose an optimum time interval $T$ over 
which an instantaneous variable like $J(t)$ or $\sigma(t)$ should be measured 
(so that $\Delta J/J$ or $\Delta \sigma/\sigma$ 
is small). BP had correctly concluded that a bacteria must employ the second strategy 
since it has to rapidly change its direction of motion based on comparison between 
its successive measurement of $\sigma(t)$. BP had estimated $T\sim 1 sec$ for {\em E. coli} 
bacteria. Recent findings \cite{natrcomments} show that bacteria has a 
very efficient mechanism for amplifying the minute signal generated by binding of 
external sugar molecules to its receptors. It has the capability of detecting 
$0.1\%$ percent change in the attractant density and that too over four orders of magnitude 
of sugar concentrations. Ref\cite{transient} has shown, that for a particular type 
of biosensor flux detection could be a superior method compared to measuring long 
time saturation properties. For our sensor, we now investigate in detail how saturation 
time of the sensor varies in response to $\rho_0$ and how it can be steered by 
choosing $\sigma_0$.

First we will discuss a simple Mean Field (MF) limit of the dynamics.    
In the MF approximation we consider the surface concentration to be uniform 
over the surface of the fiber and the volume concentration to be uniform 
through out the bulk. Let $V_0$ be the volume of the annular space and $A_0$ 
be the surface area of the fiber. At $t=0,\; N_0 = \rho_{0} V_{0}$ and later 
$N_0 =\rho_{{}_{M}} V_{0} + \sigma_{{}_{M}} A_{0}$, where $\rho_{{}_{M}}=\rho_{{}_{M}}(t)$ 
is the mean field density (denoted by subscript {\scriptsize M}) of X molecules and $\sigma_{{}_{M}}=\sigma_{{}_{M}}(t)$ is the corresponding surface 
density of the bound X molecules. In the nondimensional form we have
\begin{equation}
\tilde \rho_{{}_{M}}=1-\alpha\tilde\sigma_{{}_{M}},
\label{eq.sigma2rho}
\end{equation}
where $\tilde \rho_{{}_{M}}= \frac{\rho_{{}_{M}}}{\rho_{0}}$, 
$\tilde \sigma_{{}_{M}}= \frac{\sigma_{{}_{M}}}{\sigma_{0}}$ and  
$\alpha=\frac{\sigma_0 A_0}{\rho_0 V_0}=\frac{N_s}{N_0}$.
The bulk density $\rho_{{}_{M}}$ being homogeneous and slaved by $\sigma_{{}_{M}}$ (via Eq.\ref{eq.sigma2rho}) 
we need to consider only the equation of motion for the surface reaction, namely 
Eq.\ref{surfNonDim}.  
Substituting for $\rho_{{}_{M}}$, from Eq.\ref{eq.sigma2rho}, into Eq.\ref{surfNonDim}, and
simplifying, we get 
\begin{equation}
\frac{d\tilde{\sigma_{{}_{M}}}}{d\tau} = 
\lambda_{1}\tilde\sigma_{{}_{M}}^{2}-\lambda_{2}\tilde\sigma_{{}_{M}}+\lambda_{3},
\end{equation}
where $\lambda_{1}=\alpha\tilde\omega_b,\;
\lambda_{2}=[(1+\alpha)\tilde\omega_b + \tilde\omega_u]$ and $\lambda_{3}=\tilde\omega_b$.

Integrating this Equation we get
\begin{eqnarray}
\tau &=& \int^{\tilde\sigma_{{}_{M}}(\tau)}_{\tilde\sigma_{{}_{M}}(0)=0} \left[\frac{d\sigma_{{}_{M}}}{\lambda_{1}\sigma_{{}_{M}}^{2}-\lambda_{2}\sigma_{{}_{M}}+\lambda_{3}}\right]\\
&=&\frac{2}{i\lambda_{R}}\left[\tan^{-1}\left(\frac {\lambda_{2}}{i\lambda_{R}}\right)-
\tan^{-1}\left(\frac{\lambda_{2}-2\lambda_{1}\tilde\sigma_{{}_{M}}}{i\lambda_{R}}\right)\right] ,
\label{eq.time}
\end{eqnarray}
where $\lambda_{R}=\sqrt{\lambda_{2}^{2}-4\lambda_{1}\lambda_{3}}$.
Inverting the above equation, 
\begin{eqnarray}
\tilde\sigma_{{}_{M}} (\tau) =\frac{1}{2\lambda_{1}}\left[\lambda_{2}-\lambda_{R}\tanh\left[
\tanh^{-1}\left(\frac{\lambda_{2}}{\lambda_{R}}\right)+
\frac{\lambda_{R}\tau}{2}\right]\right].
\label{eq.sigmaT}
\end{eqnarray}
For any set of parameter values, it can be shown that $\lambda_{R}$ is always real and we also have 
$\lambda_{2}/\lambda_{R}>1$.  As a result $\tanh^{-1}\left(\lambda_{2}/\lambda_{R}\right)$ 
is always a complex number. Using the standard property $\tanh^{-1}x-\coth^{-1}x=i\pi/2$, 
we can rewrite Eq.\ref{eq.sigmaT} as
\begin{equation}
\tilde\sigma_{{}_{M}} (\tau) =\frac{1}{2\lambda_{1}}\left[\lambda_{2}-\frac{\lambda_{R}}{\tanh\left[
\tanh^{-1}\left(\frac{\lambda_{R}}{\lambda_{2}}\right)+
\frac{\lambda_{R}\tau}{2}\right]}\right].
\label{eq.sigmaTnext}
\end{equation}
This formula gives excellent fit to the numerical data (not shown here), obtained  by
integration of Eq.3 and 4, at a high value of the  diffusion constant,
$D=10^{-3} cm^2/sec$.

The steady state solution ($\tilde\sigma_{{}_{M}}^s$) can be obtained either by setting 
$\dot{\tilde\sigma}_{{}_{M}}=0$ in Eq.6 or from the $\tau \rightarrow\infty$ limit of 
Eq.\ref{eq.sigmaT}. We get 
$\tilde\sigma_{{}_{M}}^s =\frac{1}{2\lambda_{1}} \left[\lambda_{2}- \lambda_{R}\right]$.

\begin{figure}[h]
\begin{center}
\epsfxsize=7.5cm
\centerline{\epsfbox{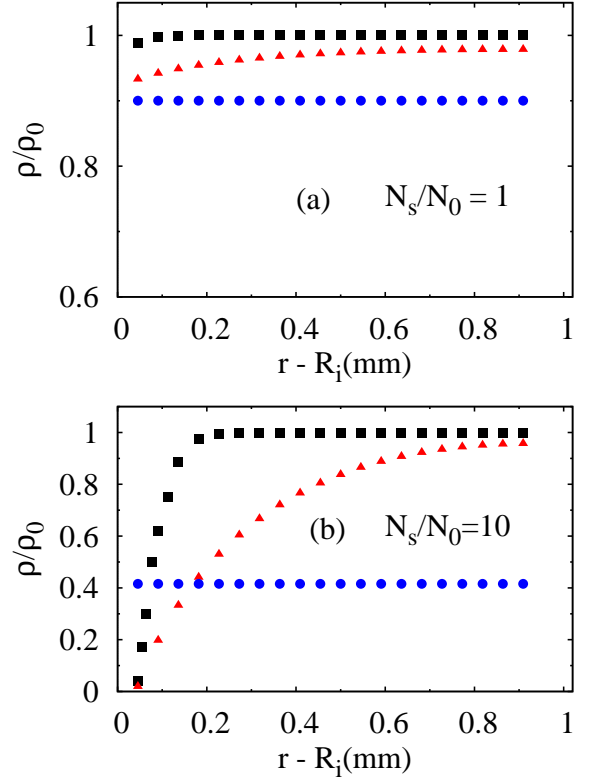}}
\caption{
Radial density profile $\rho$ as a function of $r-R_i$ in the bulk, at three
different times: right after start (square), at saturation (circle) and at
some intermediate time (triangle). (a) and (b) differ in the parameter $N_s/N_o$ 
(which is proportional to $\sigma_0$ at fixed $\rho_0$).
The values of $D$, $\omega_{b}$, $\omega_{u}$ were obtained through Fig.\ref{fig.absVsTime}.
Transition from mean field to non-mean field type density profile occurs
as we go from (a) to (b) by increasing $N_s/N_o$. 
But note that, at a fixed
$\rho_0$, the fraction of antigens (X) remaining in the bulk can be reduced
(consequently the bound proportion can be increased) by increasing
$N_s/N_o$. This is desirable for making the sensor more sensitive,
specially when $\rho_0$ is small. 
}
\label{fig.densityProfile}
\end{center}
\end{figure}
\begin{figure}[h]
\begin{center}
\epsfxsize=7.5cm
\centerline{\epsfbox{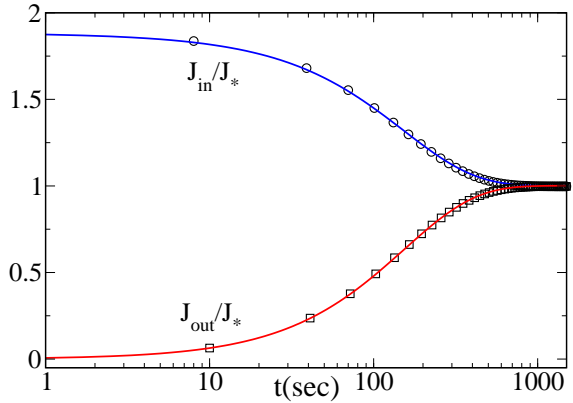}}
\caption{
Semi-log plot of binding ($J_{in}$) and unbinding ($J_{out}$) fluxes as a function of time in the mean field
regime (i.e., high $\rho_0$ and low $\sigma_0$ with $N_s/N_0=0.05$). The symbols (circles and squares) are obtained
from numerical integration of Eq.3,4 while the solid lines are the corresponding mean field 
results. Here $\rho_0=0.1 mg/ml, \sigma_0=0.01 \mu gm/mm^2$ and the values of $\omega_b$ and
$\omega_u$ are the same as those in Fig.1 (i.e., the values estimated from experimental data).
}
\label{fig:mftSimul}
\end{center}
\end{figure}

The mean field approximation will fail if diffusion is not sufficiently 
fast compared to the time scale at which surface binding reactions cause 
a depletion in the antigen concentration ($\rho$). In such a scenario the spatial 
inhomogeneity in $\rho$ (along $r$) takes a long time, comparable to the saturation
time of the sensor, to homogenize. A better understanding can be gained by comparing
the time scales of the three processes: diffusion ($t_D$), binding ($t_b$) and 
unbinding ($t_u$). We get the individual time scales from Eq.1, by comparing each 
term on the right hand side with the left hand side. 
For example, $\dot \rho\sim D\nabla ^2 \rho$ gives, by dimensional analysis, 
$t_D^{-1}\sim \frac{D}{R^2}$. Similarly, $t_b^{-1}\sim \frac{\sigma_0\omega_b}{R}$ 
and $t_u^{-1}\sim \frac{\sigma_0 \omega_b}{\rho_0R}$. Here we have assumed $R= R_o-R_i$
to be the only relevant length scale. For diffusion, this is the spatial scale of
density inhomogeneity. Now, $t_b$ and $t_u$ are the time scales over which density 
inhomogeneity are created near the fiber due to the surface reactions, while $t_D$
is the time interval during which such inhomogeneities are ironed out. Therefore, mean 
field approximation requires diffusion to be a faster process, i.e., $t_D\ll t_b,t_u$. 
These inequalities yield the criteria $\frac{\sigma_0\omega_bR}{D} \ll 1$ and 
$\frac{\sigma_0\omega_u R}{\rho_0D} \ll 1$. The first inequality suggests that
mean field approximation will be correct at high $D$ or low $\sigma_0$ values. 
We have verified these conditions numerically by looking for density inhomogeneity 
$\rho(r)$ during the transients, in the numerical solution of Eq.3,4 (see 
Fig.\ref{fig.densityProfile}). 
For example, For $N_s/N_0=1, D= 10^{-5} cm^2/sec$ the density
remain uniform through out, at all times. But when $\sigma_0$ is increased by 
choosing $N_s/N_0=10$, strongly inhomogeneous $\rho(r)$ appears (i.e., MF theory
fails). Now in addition if $D$ is hiked $\rho(r)$ becomes homogeneous 
again (graph not shown here). 
The second inequality suggests, along with high $D$ and low $\sigma_0$, we also 
need high $\rho_0$. Then only both $t_b,t_u\gg t_D$ can be satisfied. We have 
verified this condition on $\rho_0$ along with similar conditions on $\omega_b$
and $\omega_u$ resulting from the inequalities. Fig.\ref{fig:mftSimul} shows a comparison
between numerical solution of Eq.3,4 and mean field results for $\rho_0=0.1 mg/ml$ and
$\sigma_0=0.01 \mu gm/mm^2$, i.e., at high $\rho_0$ and low $\sigma_0$. At these
parameter values the influx $J_{in}$ does not go through any minima
The reason why the influx $J_{in}$ goes through a minimum in 
Fig.\ref{fig.JLowWb},\ref{fig.JHighWb} is now clear from Fig.\ref{fig.densityProfile}b, 
which shows when $\rho(r)$ becomes inhomogeneous (in the non MF case)
the $\rho(r)$ in the vicinity of the fiber undergoes a dip (triangles) before  
it becomes uniform (circles) at late times. In the mean field regime the minima
is absent because the $\rho(r)$ in the vicinity of the fiber decreases monotonically
in time as is clear from Fig.\ref{fig.densityProfile}a.

\begin{figure}[tb]
\begin{center}
\epsfxsize=7.5cm
\centerline{\epsfbox{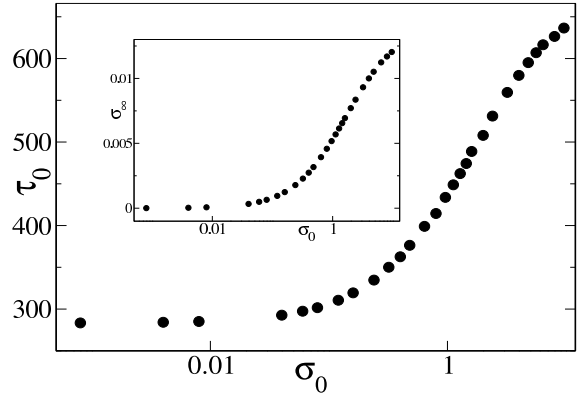}}
\caption{
{ Semi-log plot of saturation time $\tau_{0}$(sec) versus  
the density of receptors $\sigma_0(\mu g/mm^2)$. The inset shows saturated  
signal $\sigma_{\infty}$ versus $\sigma_0$. We fix $\rho_0$ at a 
very low value $0.001 mg/ml$ to test the sensitivity of the sensor. 
The values of $\omega_b$ and $\omega_u$ are those estimated before. 
Both $\tau_{0}$ and  $\sigma_{\infty}$ saturate at high values of $\sigma_0$, 
much beyond the maximum surface coverage ($0.08\mu g/mm^2)$ considered here. 
We have explored the seemingly unrealistic $\sigma_0>\sigma_0^{max}$ regime 
here because it may be possible to increase $\sigma_0$ by choosing smaller 
receptor molecules in another system.}
}   
\label{fig.saturationWtSigma}
\end{center}
\end{figure}

\begin{figure}[!htb]
{\vspace{0.5cm}}
\begin{center}
\epsfxsize=7.5cm
\centerline{\epsfbox{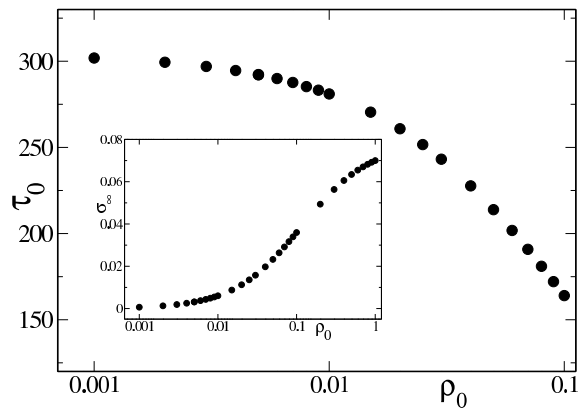}}
\caption{
Semi-log plot of saturation time $\tau_{0}$(sec) versus the density of X molecules 
$\rho_0(mg/ml)$. Here $\sigma_0$ is held fixed at the maximum surface coverage 
($0.08\mu g/mm^2$) in order to maximize the signal. 
The inset shows that even at this maximum surface coverage the saturated signal, 
$\sigma_{\infty}$ drops drastically at low $\rho_0$. 
}
\label{fig.saturationWtRho}
\end{center}
\end{figure}

We now study the general case when MF theory is invalid and thus we have to depend
on numerical integration of Eq.3,4. Our numerical curves for $\sigma$ versus time,  
shown in Fig.\ref{fig.absVsTime}, could be fit to exponential functions
like $\sigma(t)= \sigma_{\infty}(1- \exp^{-t/\tau_{0}})$, allowing us to estimate 
a saturation time scale $\tau_{0}$ and the saturated value $\sigma_{\infty}$ (plotted in 
Fig.\ref{fig.saturationWtSigma},\ref{fig.saturationWtRho}). $\tau_{0}$ and $\sigma_{\infty}$ depend on both $\sigma_{0}$ and $\rho_{0}$. 

The aim of Fig.\ref{fig.saturationWtSigma} and \ref{fig.saturationWtRho} is to identify the regimes 
where saturation time $\tau_0$ can be reduced and saturated signal $\sigma_{\infty}$ can be maximized.
For Fig.\ref{fig.saturationWtSigma}, $\rho_0$ has been held fixed 
at a low value, $0.001mg/ml$, while for Fig.\ref{fig.saturationWtRho}, $\sigma_0$ is fixed at 
$\sigma_0^{max}=0.08 \mu g/mm^2$. The insets of both the figures show that signal 
can be enhanced either by increasing  $\rho_0$ or $\sigma_0$, which result in decrease or increase
of $\tau_0$, respectively.
Of course at high $\rho_0$ a strong saturated signal can be achieved within a short saturation time, 
but the sensitivity of a sensor is tested when $\rho_0$ is small which we will focus on below. 
For low $\rho_0$ , $\sigma_0$ should
be maximum to maximize the signal, even at the cost of higher waiting time. Operating near
maximum receptor coverage is also necessary as Fig.\ref{fig.saturationWtSigma} shows that 
the nonlinear response starts to increases near this point.  But for 
moderate and high $\rho_0$, we should choose moderate
$\sigma_0$ such that $\tau_0$ is not so high and the signal is strong enough.
This may appear analogous to the conclusion of BP where with just a fraction of the 
cell area ($\sim 1/1000$) covered with receptors the steady flux could be as high 
as $J_{BP}/2$, where $J_{BP}$ is the maximum flux with the fully absorbing surface. 
But the assumption behind this derivation was that the inter-receptor distance is 
much much greater than the receptor size. In Fig.\ref{fig.saturationWtSigma}, at $1/10$-th 
of the maximum surface coverage (i.e., at $\sigma_0\sim 0.01\mu g/mm^2$) the signal $\sigma_{\infty}$ 
is much weaker compared to that at $\sigma_0^{max}$.
This again highlights the difference between our confined system and the steady state
behavior of Berg and Purcell's unbounded system.

In summary, we examined the applicability of Berg and Purcell's ideas to real
sensors. In general it turns out that a flux based sensor is more efficient 
than one which depends on long time signal. The flux in our sensor also shows 
unexpected time variation which results from competition among different time 
scales and the extended nature of our system.
Another interesting observation is that even at realistic diffusion constant, 
mean field theory works when $\rho_0$ is high and $\sigma_0$ is small. In general, 
nonspecific binding of X molecules on the fiber 
surface can cause complications but for the system we have chosen here nonspecific 
binding was verified to be negligible. Further, the surface reactions need not 
be first order, which we have assumed here. We checked that consideration of second
order binding kinetics does not give any new exotic behavior (eg, oscillations etc)
but changes the quantitative values of saturation time.\\

\end{document}